\documentclass[aps,twocolumn,floatfix]{revtex4}

\usepackage{graphicx}% Include figure files
\usepackage{dcolumn}% Align table columns on decimal point
\usepackage{bm}% bold math
\usepackage{amsmath}
\begin{document}

\newcommand{\bk}{{\bf k}}
\newcommand{\bp}{{\bf p}}
\newcommand{\bv}{{\bf v}}
\newcommand{\bq}{{\bf q}}
\newcommand{\tbq}{\tilde{\bf q}}
\newcommand{\tq}{\tilde{q}}
\newcommand{\bQ}{{\bf Q}}
\newcommand{\br}{{\bf r}}
\newcommand{\bR}{{\bf R}}
\newcommand{\bB}{{\bf B}}
\newcommand{\bA}{{\bf A}}
\newcommand{\bK}{{\bf K}}
\newcommand{\cS}{{\cal S}}
\newcommand{\vd}{{v_\Delta}}
\newcommand{\tr}{{\rm Tr}}
\newcommand{\kslash}{\not\!k}
\newcommand{\qslash}{\not\!q}
\newcommand{\pslash}{\not\!p}
\newcommand{\rslash}{\not\!r}
\newcommand{\bs}{{\bar\sigma}}

\title{4D-XY quantum criticality in a doped Mott insulator}

\author{M. Franz and A.P. Iyengar}
\affiliation{Department of Physics and Astronomy,
University of British Columbia, Vancouver, BC, Canada V6T 1Z1}
\date{\today}

\begin{abstract}
A new phenomenology is proposed for the superfluid density $\rho_s$ of strongly 
underdoped cuprate superconductors based on recent data for ultra-clean  single
crystals of YBa$_2$Cu$_3$O$_{7-x}$. We show that the 
puzzling departure from Uemura scaling and 
the decline of the slope as the $T_c = 0$ quantum critical point is approached
can be understood in terms of the renormalization of quasiparticle effective 
charge by quantum fluctuations of the superconducting
phase. We then employ (3+1)-dimensional XY model 
to calculate, within particular approximations,  the renormalization of 
$\rho_s$ and its slope, explain the new phenomenology, and
predict its eventual demise close to the QCP.
\end{abstract}
\maketitle

The manner in which superconductivity 
in high-$T_c$ cuprates gives way to Mott insulating behavior is 
a longstanding puzzle of fundamental importance.
The anomalous behavior is revealed most strikingly in studies of the doping $(x)$ 
and temperature $(T)$ dependence of the superfluid density $\rho_s$ 
\cite{hardy1,wen1}.
As the doping is reduced, both $\rho_s$ and the critical temperature
$T_c$ decline while the maximum superconducting 
gap $\Delta_0$ increases. This dichotomy, along with the empirical Uemura
relation $T_c\propto\rho_s$ \cite{uemura1}, led to suggestions by numerous 
authors \cite{emery1,randeria1,fm1,balents1,ft1} that the superconducting 
transition
in the underdoped cuprates is a {\em phase-disordering} transition and the 
pseudogap state \cite{timusk1} above $T_c$ should be thought of as a 
``phase-disordered'' $d$-wave superconductor. Experiments indeed show 
evidence for magnetic vortices \cite{ong1}, strong fluctuation 
diamagnetism \cite{ong2}, and fermionic nodal
quasiparticles \cite{sutherland1} in the pseudogap state of
high-$T_c$ compounds. In addition, there is credible evidence that the 
superconducting transition in many compounds is in the 3D-XY universality
class \cite{salamon1,hardy2,meingast1,harlingen1} with a wide critical region, exactly as one would expect 
near a phase-disordering transition.

The focus of  the present  Letter is to obtain a theoretical understanding of 
the behavior of $\rho_s$ in {\em strongly} underdoped cuprates. 
Theoretical efforts to date have overwhelmingly addressed
the phenomenology summarized by Lee and Wen \cite{wen1}.
However, recent experiments performed in unprecedented proximity to the 
Mott insulator in ultraclean single crystals of YBa$_2$Cu$_3$O$_{7-x}$ 
(YBCO) \cite{tami1,liang1,broun1} and in high quality
films \cite{lem1}  are quietly overturning the old paradigm: 
%revealing several key new features of the strongly underdoped regime.  
($i$) Unlike optimally doped and weakly underdoped samples, the strongly 
underdoped data show
no visible 3D-XY critical region in samples with $T_c\lesssim 25$K; rather,
the approach to $T_c$ is mean-field like. ($ii$) 
Measurements indicate a relationship between the critical temperature
and the $T=0$ superfluid density of the form
\begin{equation}
T_c\propto\rho_s^\gamma,\ \  {\rm with} \ \ \gamma\simeq 0.4-0.7,
\label{ue1}
\end{equation}
a significant departure from the Uemura scaling.  
($iii$) The overall doping and temperature dependence
can be parametrized as
\begin{equation}
\rho_s(x,T)\simeq Ay^2-By(k_BT),
\label{rho1}
\end{equation}
where $y=T_c(x)/T_c^{\rm max}$ is a measure of the doping $x$,
$T_c^{\rm max}=93$K is
the maximum critical temperature for YBCO,  $A\simeq 66$meV, and
$B\simeq 9.5$.
The demise of superconducting order described by Eq.\ (\ref{rho1})
signals a profound departure from predictions of 
RVB-type theories and earlier parametrizations of  $\rho_s(x,T)$
\cite{wen1,millis98}.  In what follows, we demonstrate how this 
new phenomenology follows simply and elegantly from an effective theory 
describing a phase-fluctuating $d$-wave superconductor.
\begin{figure}
\includegraphics[width = 6.0cm]{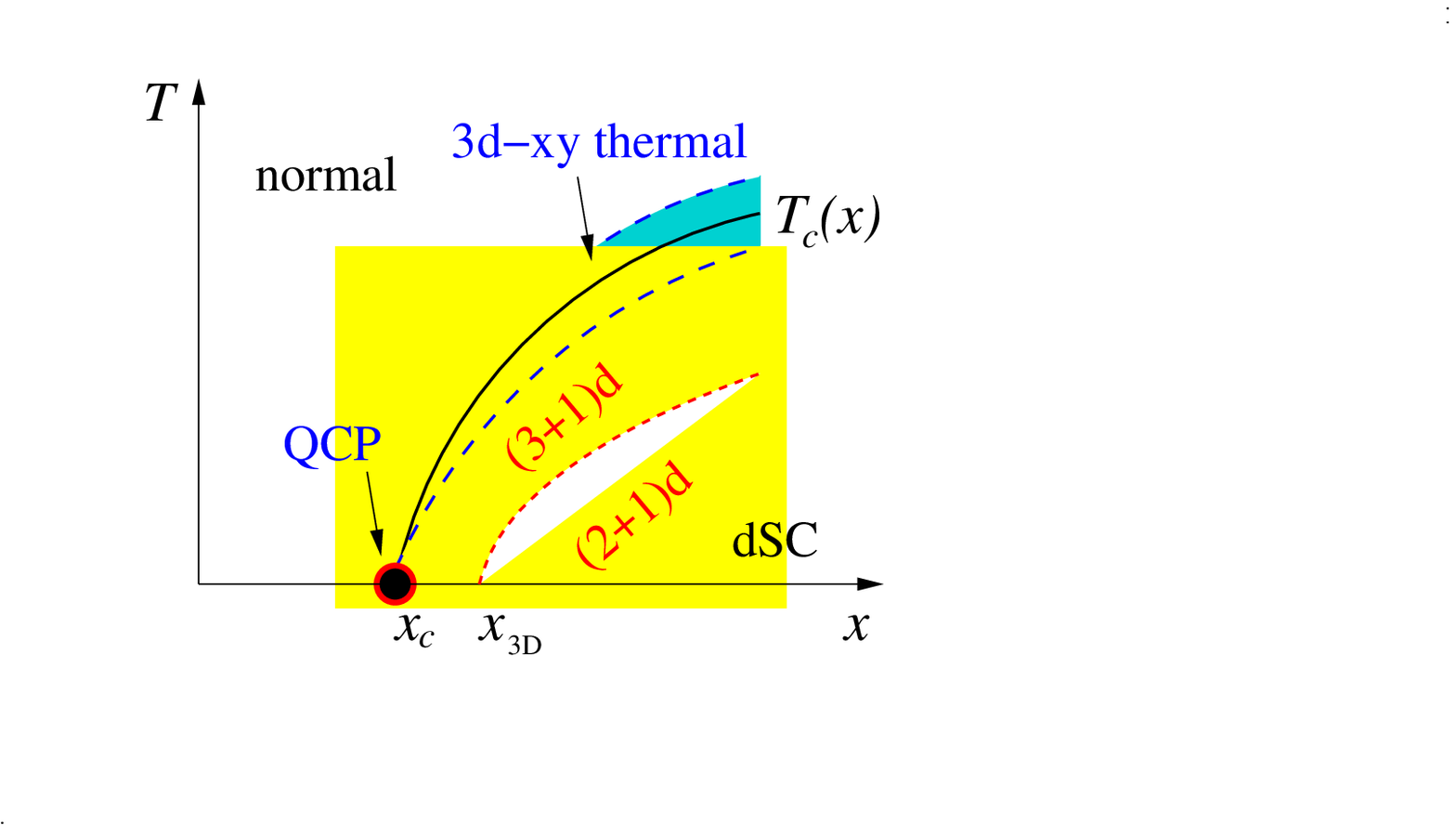}
\caption{Behavior expected in the vicinity of the $(3+z)$D-XY  quantum 
critical point in the doping--temperature plane. Solid line $T_c(x)$ represents
the superconducting phase transition; dashed lines are crossovers. 
}
\label{fig1}
\end{figure}

The absence of any visible 3D-XY fluctuation region in the data is surprising 
since one would naively expect phase fluctuation effects to become more 
pronounced in the underdoped region as the system approaches the Mott insulator.  
Upon closer inspection, however, one finds that the observed behavior 
is entirely consistent with the behavior of a system approaching a quantum 
critical point \cite{millis1,sachdev_book}. Indeed, the $T_c(x)$
line in Fig.\ \ref{fig1} must terminate at a quantum critical point 
(QCP), which
by continuity must be in the universality class of the $(3+z)$-dimensional XY model, the imaginary time $\tau$ providing the extra $z$ 
dimensions ($z\geq 1$ being the dynamical critical exponent.) Since $D=4$ is the
upper critical dimension for XY-type models, our QCP sits either right at $(z=1)$
or above $(z>1)$ its upper critical dimension, and thus we expect 
{\em mean field} critical
behavior, possibly with unimportant logarithmic corrections if $z=1$  
\cite{millis1,sachdev_book}. As indicated in Fig.\ \ref{fig1}, when crossing
the finite-temperature transition close to the QCP one still encounters the 
classical fluctuation region, but its width is now much reduced, and it is
likely invisible in experiments.

There is a simple consistency check for the above scenario. 
If the underdoped region
is indeed controlled by the $(3+z)$D-XY point then  there exists a simple scaling
relation between $T_c(x)$ and $\rho_s(x,0)$ which reads \cite{millis1,herbut1}
\begin{equation}
T_c\propto \rho_s^{z/(d-2+z)}.
\label{ue2}
\end{equation}
For $d=2$ we recover the Uemura scaling, irrespective of the value of $z$. In
$d=3$, as it appears to be the case in YBCO, we get 
$T_c\propto \rho_s^{z/(1+z)}$, consistent with the experimental observation 
of Eq.\ (\ref{ue1}), if $1\leq z \leq 2$.

It would thus appear that general arguments based on the proximity of the
underdoped cuprates to a putative $(3+z)$D-XY QCP naturally explain items
($i$) and ($ii$) in the foregoing list. The scaling analysis leading to 
Eq.\ (\ref{ue2}), however, only holds when the ``bare'' parameters of the theory 
describing the critical degrees of freedom exhibit no significant temperature 
dependence. In cuprates, quasiparticles in the vicinity of $d$-wave nodes give 
rise to a linear
temperature dependence in $\rho_s$ which is likely to modify the scaling.
In order to address this issue and item ($iii$) one needs to go beyond the general 
scaling arguments and consider a specific model. In the rest of the paper, 
we formulate and study a particular version of the quantum XY model.
We show that, when nodal quasiparticles are included through the effective 
parameters of this model, it leads to a phenomenology that is consistent 
with the data.

The simplest model showing the XY-type critical behavior 
is given by the Hamiltonian
\begin{equation}
H_{\rm XY}={1\over 2}\sum_{ij}\hat{n}_iV_{ij}\hat{n}_j
-\frac{1}{2}
\sum_{ij}J_{ij}\cos(\hat{\varphi}_i-\hat{\varphi}_j).
\label{hxy}
\end{equation}
Here $\hat{n}_i$ and $\hat{\varphi}_i$ are the number and phase operators
representing 
Cooper pairs on site $\br_i$ of a cubic lattice and are quantum mechanically
conjugate variables, $[\hat{n}_i,\hat{\varphi}_j]=i\delta_{ij}$.
The sites $\br_i$ do not necessarily represent individual 
Cu atoms; rather one should think in terms of ``coarse grained'' lattice model
valid at long lengthscales where microscopic details no longer matter.
Classical \cite{stroud1,carlson1} and quantum \cite{paramekanti1,herbut1} versions
of the XY model have been employed previously to study phase fluctuations
in the cuprates.

The first term in $H_{\rm XY}$ describes interactions
between Cooper pairs; we take
\begin{equation}
V_{ij}=U\delta_{ij}+(1-\delta_{ij}){e^2\over |\br_i-\br_j|}.
\label{vij}
\end{equation}
The second term in $H_{\rm XY}$ represents the Josephson tunneling of
pairs between the sites; $J_{ij}= J$ for nearest neighbors along the $a$ and $b$ 
directions, $J'$ along $c$, and $0$ otherwise. 
In the absence of interactions, $J$ clearly must be identified as the
physical superfluid density. We thus take
\begin{equation}
J=J_0-\alpha T
\label{j}
\end{equation}
with $\alpha=(2\ln{2}/\pi)v_F/v_\Delta$, as in a BCS $d$-wave superconductor.
The $T$-linear term describes suppression of the mean-field superfluid
stiffness by nodal excitations.

On a qualitative level, the physics of the quantum XY model (\ref{hxy}) can be 
understood in terms of the competition between the fluctuations in the local 
phase 
$\hat\varphi_i$ and charge $\hat n_i$. These are constrained by the uncertainty
relation $\Delta\varphi_i\cdot\Delta n_i\geq 1$ which implies that interactions,
which tend to localize charge, also necessarily promote phase fluctuations, 
which then erode the superfluid density. Ultimately, for sufficiently strong 
$V_{ij}$,
a superconductor-insulator (SI) transition takes place. Charge becomes
localized,
and an insulating pair Wigner crystal is formed \cite{zlatko1}. 
The latter can be 
viewed as a superconductor with completely disordered phase. In the rest of 
the paper we assume that the strength of interactions 
{\em increases} with underdoping and use $V_{ij}$ to tune our model across
the SI transition. 

To gain quantitative insight into the behavior of $\rho_s$ in the XY model
(\ref{hxy}), we employ two complementary approaches: the self-consistent
harmonic approximation (SCHA) \cite{stroud1,paramekanti1}, valid for weak
interactions, and an expansion in the small order parameter 
\cite{doniach1,fisher1}, valid for strong interactions in the vicinity of the 
SI transition. 

In the SCHA one replaces $H_{\rm XY}$ by the ``trial'' harmonic Hamiltonian
\begin{equation}
H_{\rm har}={1\over 2}\sum_{ij}\hat{n}_iV_{ij}\hat{n}_j+
{1\over 2}\sum_{\langle ij\rangle}
K_{ij}(\hat{\varphi}_i-\hat{\varphi}_j)^2.
\end{equation}
The constants $K_{ij}=K\:(K')$ are identified as the renormalized $ab$-plane
($c$-axis)
superfluid densities \cite{stroud1}, and are determined from the 
requirement that 
$E_{\rm har}\equiv\langle H_{\rm XY} \rangle_{\rm har}$ be minimal.
%This variational principle can be extended to
%$T>0$ case using the Gibbs-Bogoliubov inequality $F\leq F_{\rm har}+
%\langle H-H_{\rm har}\rangle_{\rm har}$.
%%The cuprates are characterized by small values of the anisotropy ratio $\eta=J'/J$.  
%%To get some feeling for the impact of anisotropy on the phase fluctuations
%%we first discuss two limiting cases, $\eta=1$ (isotropic 3D superconductor) 
%%and $\eta=0$ (decoupled 2D layers), which can be solved analytically. We
%%also provide numerical results for intermediate anisotropy.  
The cuprates are characterized by small values of the anisotropy ratio $\eta=J'/J$.
Anisotropy essentially interpolates between the cases $d=2$ and $d=3$ and 
thus profoundly affects the approach to the {\em thermal} phase transition. 
However, for the quantum phase transition the effect is very weak. We illustrate this 
point below by solving the limiting cases $\eta=1$ (isotropic 3D superconductor) 
and $\eta=0$ (decoupled 2D layers) analytically and the intermediate case 
$0 < \eta < 1$ numerically.

The trial Hamiltonian
$H_{\rm har}$ is quadratic in $\hat{n}_i$ and $\hat{\varphi}_j$ 
and can thus be easily diagonalized,
\begin{equation}
H_{\rm har}=\sum_\bq \hbar\omega_\bq(a^\dagger_\bq a^{}_\bq+{1\over 2}),
\ \ \ \ \hbar\omega_\bq=2\sqrt{KZ_\bq V_\bq},
\label{har}
\end{equation}
where $V_\bq$ is a Fourier transform of $V_{ij}$ and 
$Z_\bq=\sum_{\mu=1}^d\sin^2(q_\mu/2)$, $d=2,3$.
For short range interactions $V_\bq\to$ {\sl const} as $q\to 0$; we have
$\omega_\bq\sim q$, i.e. an acoustic phase mode.
For Coulomb interactions, $V_\bq\sim 1/q^2 $ as $q\to 0$; we have
$\omega_\bq\to \omega_{\rm pl}$, i.e. a gapped plasma mode.
Simple power counting then shows that at low $T$ the contribution from the
phase mode to the superfluid density is
\begin{equation}
\delta\rho_s^{\rm ph}\sim\biggl\{
\begin{array}{ll}
T^{d+1}, \ \ \ \ \ &{\rm short\ range\ interaction} \\
e^{-\omega_{\rm pl}/T},& {\rm Coulomb\ interaction}
\end{array}
.
\end{equation}
In either case the low-$T$ behavior of $\rho_s$ will be dominated by 
the quasiparticle contribution included via Eq.\ (\ref{j}). However, as 
we demonstrate 
below {\em quantum fluctuations} of the phase lead to strong renormalizations 
of both the $T=0$ amplitude of $\rho_s$ and the slope $\alpha$.

\begin{figure}
\includegraphics[width = 7.0cm]{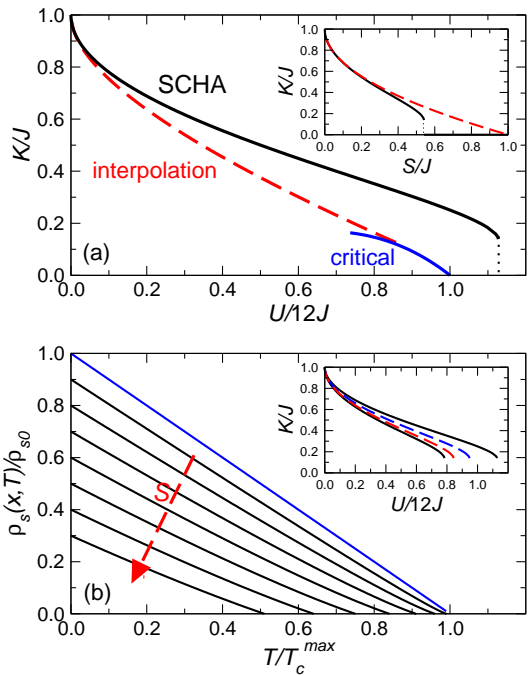}
\caption{Panel (a) shows SCHA result, 
the critical theory Eq.\ (\ref{rhos-cr}), and the interpolation discussed in 
the text (dashed line) for the case $d=3$ with short-ranged interactions. 
Inset: solid line is the exact solution of Eq.\ 
(\ref{Ka}) and the dashed line is its approximation Eq.\ (\ref{Kb}). 
Panel (b) displays $\rho_s(x,T)$ from Eq.\ (\ref{Kb})
for several values of $S$. Inset illustrates the effect of 
anisotropy: the curves (bottom to top) correspond to $\eta=0.0, 0.3, 0.6, 1.0$.
Solid lines are solutions of Eq. (\ref{Ka}) for $d=2,3$ and short range 
interaction (we find $S/U \approx 0.040$ for $d=3$ and $S/U \approx 0.058$ for $d=2$). 
Dashed lines are numerical solutions of the corresponding anisotropic 
equations. 
}
\label{fig2}
\end{figure}

Using $\langle\cos(\hat{\varphi}_i-\hat{\varphi}_j)\rangle_{\rm har}=
\exp{[-{1\over 2}\langle(\hat{\varphi}_i-\hat{\varphi}_j)^2 \rangle_{\rm har}]}$,
an identity valid for harmonic Hamiltonians, we obtain
\begin{equation}
E_{\rm har}=\langle H_{\rm XY} \rangle_{\rm har}=
\sqrt{KS}-Je^{-\sqrt{S/K}},
\end{equation}
with the parameter $\sqrt{S}=(2dN)^{-1}\sum_\bq\sqrt{V_\bq Z_\bq}$ 
describing the aggregate strength of interactions.
Minimizing $E_{\rm har}$ with respect to $K$ we find
\begin{subequations}
\begin{eqnarray}
\label{Ka}
K&=&Je^{-\sqrt{S/K}} \\
&\simeq&  J(1-\sqrt{S/J}),
\label{Kb}
\end{eqnarray}
\end{subequations}
where the last expression approximates the exact solution over much of the 
regime of interest (see inset to Fig.\ \ref{fig2}a.)
The exact solution of Eq.\ (\ref{Ka})  exhibits a first
order transition at $S\approx 0.541\:J$. This is an artifact of the SCHA; 
close to the transition
phase fluctuations become of the order of $\pi/2$ and
the harmonic Hamiltonian (\ref{har}) is no longer a legitimate approximation
to $H_{\rm XY}$. Below we devise an interpolation formula based on
Eq.\  (\ref{Kb}) that will represent an acceptable solution everywhere except 
very close to the SI transition.

To obtain the leading temperature dependence, we substitute
$J=J_0-\alpha T$ into Eq.\  (\ref{Kb}) and expand to leading order in $T$:
\begin{equation}
\rho_s(x,T)\simeq J_0{\left(1-\sqrt{S/J_0}\right)}
-\alpha T {{\left(1-{\small 1\over 2}\sqrt{S/J_0}\right)}}.
\label{rhos-t}
\end{equation}
This is our main result. As expected, 
both the amplitude and the slope are 
reduced by quantum fluctuations. Crucially, we observe that the $T=0$ 
amplitude decays {\em faster} than the slope.
In particular, for $\sqrt{S/J_0}$ not too large the above 
expression is consistent with the experimentally observed behavior
Eq.\ (\ref{rho1}) if we identify $y\simeq (1-{1\over 2}\sqrt{S/J_0})$.
In the language of Ref.\ \cite{millis98} the parameter $y$ 
can be interpreted as the quasiparticle charge renormalization factor.

If we follow Lee and Wen \cite{wen1} and determine $T_c$ as the 
temperature at which the
superfluid stiffness vanishes, then Eq.\ (\ref{rhos-t}) implies, to leading
order, that  $\rho_s(x,0)\sim T_c^2$, in agreement with the empirical relation
Eq.\ (\ref{ue1}). 
The agreement with the scaling result (\ref{ue2}) is, however, 
entirely coincidental since our description 
of the superfluid density in the SCHA involves an interplay between nodal 
quasiparticles  and {\em noncritical} quantum phase fluctuations.

For arbitrary anisotropy $0<\eta<1$, the SCHA yields a pair of equations for 
$K$ and $K'$ with structure similar to Eq.\ (\ref{Ka}). These are easily
solved numerically \cite{iyengar1}, and we give some representative results 
in the inset to Fig. \ref{fig2}b. Inspection of this data reveals that all 
the characteristic features of the $d=2,3$ limiting cases remain in place for 
general anisotropy. For realistic anisotropies 
$\eta=10^{-2}-10^{-3}$ the results become practically indistinguishable from 
the $d=2$ case. 

In the regime of {\em strong} fluctuations it is useful to consider the grand 
canonical
partition function for $H_{\rm XY}$ expressed in the path-integral representation
as a trace over bose field $\varphi_i(\tau)$, $Z=\int{\cal D}\varphi
\exp(-\cS/\hbar)$,
with the action
\begin{equation}
\cS={1\over 2}\int_0^\beta d\tau\sum_{ij}\left[\dot{\varphi}_iV^{-1}_{ij}
\dot{\varphi}_j- J_{ij}\cos({\varphi}_i-{\varphi}_j)\right].
\label{sxy}
\end{equation}
Following Refs.\ \onlinecite{doniach1} and\ \onlinecite{fisher1},
we introduce an auxiliary complex field $\psi_i(\tau)$ to decouple, via
the familiar Hubbard-Stratonovich transformation, the cosine term in the above
action. For short range interaction the decoupled action is local in the 
$\varphi_i(\tau)$ field and the ${\cal D}\varphi$ functional integral can be 
performed {\em exactly}, to any order in powers of $\psi$ and its derivatives. 
The field $\psi$ assumes the role of the
order parameter of the SI transition. Keeping only terms up to $|\psi|^4$ and 
replacing the spatial lattice by the continuum yields the desired 
field-theoretic
representation $Z=\int{\cal D}\psi\exp(-\cS_{\rm eff}/\hbar)$, where
\begin{equation}
\cS_{\rm eff}=\int d\tau d^d x \left\{r|\psi|^2+{u\over 2}|\psi|^4+
{1\over 2}|\nabla\psi|^2 +{1\over 2c^2}|\partial_\tau\psi|^2\right\}
\nonumber
\label{seff}
\end{equation}
and $r=(d/a_0^2)(1-4dJ/U)$, $u=(7da_0^{d-4}/8)J^2(4d/U)^3$, and 
$c^2=(4da_0^2/3)(U/4d)^3/J$, with $a_0$ the lattice spacing.

The above action $\cS_{\rm eff}$ predicts a second order SI transition when $r$ 
changes sign, i.e.\ when $U=U_c=4dJ$. Anisotropy again interpolates smoothly 
between the limiting cases \cite{iyengar1} and gives $U_c=4(2+\eta)J$. 
In $(3+1)$ dimensions we expect mean field
theory to work near this transition. In particular, the superfluid density will
be given by the saddle-point value of the order parameter  $|\psi_0|^2
=-r/u$ which yields
\begin{equation}
\rho_s(x,T)={8\over 7}\left({U\over 12J}\right)^2(J-U/12)
\label{rhos-cr}
\end{equation}
with $J$ given by Eq.\ (\ref{j}) and doping parametrized by $U$.
The Coulomb interaction can also be incorporated in  $\cS_{\rm eff}$ by 
introducing a gauge field, but the analysis near the critical point of the 
resulting action is more involved \cite{fisher1} and beyond the scope of 
this Letter.

The main panel of Fig.\ \ref{fig2}(a) combines Eq.\ (\ref{rhos-cr}) with 
the SCHA result adapted to the case of a short range interaction, for which
$S=0.48\,(U/12)$. The actual solution must interpolate smoothly between SCHA at 
small $\sqrt{S/J}$ and critical theory  near the transition. 
The dashed line represents an empirical extension of 
Eq.\ (\ref{Kb}) to  $K=J[1-\sqrt{S/J}-\lambda(S/J)]$ with
$\lambda=0.625$, which we expect to be very close to the exact 
solution, as can be verified by quantum Monte Carlo or a similar
technique. This interpolation
still exhibits the leading behavior of Eq.\ (\ref{rhos-t}), consistent with
experimental data \cite{tami1,liang1,broun1,lem1} as summarized by Eq.\ 
(\ref{rho1}).

Our results thus lend further support to the picture of underdoped 
cuprates as superconductors with a large pairing gap scale but superfluid 
stiffness that is severely suppressed by Mott physics. In our approach, the 
latter is modeled by the charging energy terms in the 
XY Hamiltonian (\ref{hxy}) which significantly renormalize both the $T=0$ 
amplitude of the superfluid density $\rho_s(x,T)$ and the quasiparticle 
effective charge reflected by the slope of its $T$-linear term. 
A key new observation of this work is that the systematics of these 
renormalizations matches that found in underdoped cuprates.
In particular the suppression of the amplitude is {\em faster} than that 
of the slope, in agreement with the experimental 
data \cite{tami1,liang1,broun1,lem1} summarized in Eq.\ (\ref{rho1}). 
As illustrated in Fig.\ \ref{fig2}, this 
behavior persists over a wide range of interaction strengths. Close to the
SI transition, the critical theory  $\cS_{\rm eff}$ takes over. In this regime  
$\rho_s(x,T)$ is given by Eq.\ (\ref{rhos-cr}), which implies that the slope of 
the $T$-linear term stops decreasing and in fact begins to increase. 
Thus, our model offers a testable prediction that very close to the SI 
transition the phenomenology of Eq.\ (\ref{rho1}) will ultimately break down.
While the quantitative details of these predictions depend on the specifics 
of our model and the approximations employed, the general features are 
controlled by the symmetry and dimensionality and should be robust.

The authors are indebted to I. Affleck, 
P.W. Anderson, A.J. Berlinsky, D.M. Broun, D.A. Bonn,
W.N. Hardy, I.F. Herbut, A.J. Millis, J. Moore, A. Paramekanti, S. Sachdev, and 
Z. Te\v{s}anovi\'c for stimulating discussions and correspondence. This work 
was supported by NSERC, CIAR and the A.P. Sloan Foundation.

\end{document}